\documentclass[conference]{IEEEtran}

\usepackage{amsmath}

\usepackage{amssymb}
\usepackage{amsfonts}
\usepackage{hyperref}
\usepackage{algorithmic} 
\usepackage{graphicx}
\usepackage{textcomp}
\usepackage{xcolor}
\usepackage{xspace}
\usepackage[ruled,linesnumbered]{algorithm2e}
\usepackage{multirow}
\usepackage{booktabs}
\usepackage{soul}
\usepackage{balance}
\usepackage{tcolorbox}
\usepackage{array}
\usepackage{multirow}
\usepackage{enumitem}
\usepackage{adjustbox}
\usepackage{cite}
\def\BibTeX{{\rm B\kern-.05em{\sc i\kern-.025em b}\kern-.08em
    T\kern-.1667em\lower.7ex\hbox{E}\kern-.125emX}}

\newcommand\tool{\textsc{DeepMetis}\xspace} 
\newcommand\DC{\textit{DeepCrime}\xspace} 
\newcommand\genmodel{input generation model}

\newboolean{showcomments}
\setboolean{showcomments}{true}         
\ifthenelse{\boolean{showcomments}}
  {\newcommand{\nb}[2]{
  \fbox{\bfseries\sffamily\scriptsize#1}
     {\sf\small$\blacktriangleright$\textit{\textcolor{red}{#2}}$\blacktriangleleft$}
   }
  }
  {\newcommand{\nb}[2]{}
   
  }

\def\tcc{\leavevmode\rlap{\hbox to \hsize{\color{gray!35}\leaders\hrule height .8\baselineskip depth .5ex\hfill}}}

\begin{document}

\title{\tool: Augmenting a Deep Learning Test Set to Increase its Mutation Score}


\author{\IEEEauthorblockN{Vincenzo Riccio, Nargiz Humbatova, Gunel Jahangirova and Paolo Tonella}
\IEEEauthorblockA{Universit{\`a} della Svizzera Italiana\\
Lugano, Switzerland\\
Email: name.surname@usi.ch}
}

\maketitle

\IEEEoverridecommandlockouts

\IEEEpubid{\begin{minipage}{\textwidth}\ \\[45pt]
  \copyright 2021 IEEE.\\
  This is the authors’ version of the work. It is posted here for your personal\\ use. Not for redistribution. The definitive Version of Record is published in\\ Proceedings of ASE ’21, Nov 15–19, 2021, Virtual, Australia\\
\end{minipage}} 

\begin{abstract}
Deep Learning (DL) components are routinely integrated into software systems that need to perform complex tasks such as image or natural language processing. The adequacy of the test data used to test such systems can be assessed by their ability to expose artificially injected faults (mutations) that simulate real DL faults.

In this paper, we describe an approach to automatically generate new test inputs that can be used to augment the existing test set so that its capability to detect DL mutations increases. Our tool \tool implements a search based input generation strategy. To account for the non-determinism of the training and the mutation processes, our fitness function involves multiple instances of the DL model under test. Experimental results show that \tool is effective at augmenting the given test set, increasing its capability to detect mutants by 63\% on average. A leave-one-out experiment shows that the augmented test set is capable of exposing unseen mutants, which simulate the occurrence of yet undetected faults.
\end{abstract}


\section{Introduction}
\label{introduction}

Deep Learning (DL) based software is widespread and has been successfully applied to complex tasks such as image processing and speech recognition. Systems including DL components are also employed in safety and business-critical domains, e.g. autonomous driving and financial trading. 
DL systems possess the human-like ability to learn how to perform a task from experience, i.e., the inputs seen during training~\cite{Manning-IIR-2008}, but such ability comes with the possibility to make errors when presented with new inputs. Therefore, it is crucial for DL software developers and manufacturers to assess to what extent these systems can be trusted in response to real-world inputs, as they could face scenarios that might be not sufficiently represented in the data from which they have learned. 

Traditional test adequacy criteria, like code coverage, fail to determine whether DL systems are adequately exercised by a test set since most of the DL systems' behaviour depends on their training data, not the code. Recent research defined ad-hoc white-box adequacy metrics, based on DL software's internal architecture, e.g., neuron~\cite{PeiCYJ17,GuoJZCS18,TianPSB18,XieISSTA19} or surprise coverage~\cite{KimFY19}. A limitation of these approaches is that their output cannot be directly associated with a root cause of a DL system's failure, i.e., a DL fault~\cite{Harel-CanadaWGG20}.


\IEEEpubidadjcol

On the other hand, mutation testing approaches evaluate a test set against faults that are artificially injected into the system under test. So, the inability of a test set to expose injected faults (\textit{kill mutants} in the mutation testing jargon) can be interpreted as its inability to properly exercise the mutated code~\cite{JiaTSE11}. The tool \DC generates mutants of DL systems by injecting artificial faults that resemble those described in the taxonomy of real DL faults by Humbatova et al.~\cite{HumbatovaICSE20}. In this way, it addresses the challenge of simulating real-world DL faults~\cite{Zhang20}. Hence, a DL test set that cannot kill a mutant generated by \DC is also unlikely to expose any real fault similar to the one injected by \DC, in case such a fault affected the DL system under test. In such a situation, the test set should be augmented with additional tests that target the undetected fault.

In this paper, we introduce a novel and automated way to augment existing test sets with inputs that kill mutants generated by \DC. Our goal is to increase the mutation score of a test set by generating new inputs that kill the mutants not killed by the original test set. To this aim, we propose \tool, a search-based test generator for DL systems that uses mutation adequacy as guidance. 
Intuitively, 
a mutant is killed if the correct behaviour is observed for a DL model under test, while a misbehaviour is observed on its mutated version. 
However, mutation testing approaches should take into account the stochastic nature of DL (in particular, of its training process) and of mutation generation (some DL mutations are non-deterministic) to properly measure the test set's ability to discriminate the original system from the artificially generated faulty versions~\cite{RiccioEMSE20}. 
In fact, observing a drop in accuracy between the original and the mutated model is not enough to conclude that the mutant is killed since such a drop might be due to random fluctuations of accuracy associated with the non-determinism of the training and the mutation process. The mutation killing criterion proposed by Jahangirova and Tonella~\cite{JahangirovaICST20} addresses this DL-specific challenge by evaluating a test set on multiple re-trained instances of the same model and applying statistical tests. \tool adopts the same non-deterministic view on DL systems, and correspondingly, its generation process is guided by multiple instances of the model being mutated. Recently, DL-specific mutation operators have been used for different tasks, such as program repair~\cite{Sohn:2019}, adversarial inputs detection~\cite{Wang:2019}, generation of adversarial code snippets~\cite{PourICST21}, and calculation of optimal oracles for autonomous  vehicles~\cite{QualityMetrics2021}, but no approach leveraged them to generate new inputs which augment an inadequate test set.

We evaluated \tool on both a classification problem and a regression problem, using mutation operators provided by \DC. Results show that \tool is effective at generating inputs that improve a test set in terms of its mutation killing ability. 
We also conducted a leave-one-out experiment to simulate a practical usage scenario where an undetected fault affecting the DL system is unknown. 
In this experiment setting, one mutant produced by \DC is taken apart, while test augmentation is performed by \tool based only on the remaining mutants. In this way, the left-out mutants simulate a yet unknown fault. Results show that left out mutants can be killed by the augmented test set on average 82\% of the time.


\section{Background}
\label{background}
\subsection{Mutation Testing of DL Systems} \label{sec:mutation_testing}

Mutation testing is a technique that injects artificial faults into a system under test guided by the assumption that the ability to expose such artificial faults translates into the ability to expose also real faults. In traditional software systems, the main decision logic of a program is implemented in its source code, and synthetic faults are introduced by applying small syntactic changes to the source code. In contrast, the behaviour of a DL system is determined not only by the source code but also by its training data, the structure of its neural networks or the tuning of various hyperparameters. As syntactic code changes are not sufficient to achieve realistic fault injection, DL mutation operators have a different nature~\cite{JahangirovaICST20}.

DeepMutation~\cite{Ma:2018} and MuNN~\cite{Shen:2018}  were the first works to recognise the need for mutation operators tailored specifically to DL systems. In DeepMutation (later extended into a tool called \textit{DeepMutation++} \cite{Ma:2019}), the authors propose a set of operators of two distinct categories: \textit{source level} and \textit{model level} operators. Source level operators apply changes to training data or model structure before training is performed, while model level operators alter weights, biases or the structure of an already trained model. Model level mutation operators tend to be less costly as, unlike source-level operators, they do not require re-training. Mutation operators proposed in MuNN \cite{Shen:2018} solely belong to the latter category.


Jahangirova \& Tonella \cite{JahangirovaICST20} performed an extensive empirical evaluation of the mutation operators proposed in DeepMutation++ and MuNN and investigated the configuration space of their parameters. For example, for a mutation operator that imitates training with corrupted data by changing the labels of training inputs to incorrect ones, the parameter would be the percentage of mutated labels. 
According to their results, the choice of the parameter values affects the impact of the mutation to a major extent.

Moreover, the authors propose a novel mutation killing criterion, which takes into account the stochastic nature of DL systems. Their definition requires multiple re-trainings of both the original program and the mutant to obtain $n$ distinct model \textit{instances} of each ($n = 20$ in their experiments).
Then, they measure whether the difference between 20 accuracies (or any other quality metrics) obtained on original vs mutated model instances is statistically significant (\textit{p\_value} $< 0.05$) and whether the effect size is not ``negligible''. 
If these conditions hold, the mutation is considered \textit{killed}. 

\subsection{DeepCrime}\label{sec:deepcrime}

\DC \cite{DeepCrime} is a mutation testing tool designed for automated seeding of artificial faults (mutations) into DL systems. Its main difference from DeepMutation++ is that \DC is based on a set of mutation operators derived from \textit{real faults}. In \DC the authors propose 35 and implement 24 \textit{source level} mutation operators that target different aspects of the development and training of DL systems. This set of operators was extracted from an existing taxonomy of real faults in deep learning systems \cite{HumbatovaICSE20} and was complemented with the issues found in the replication packages for the studies by Islam et al. \cite{Islam:2019} and Zhang et al. \cite{Zhang:2018}. To establish whether a mutation is killed or not, \DC incorporates the notion of statistical killing proposed by Jahangirova \& Tonella \cite{JahangirovaICST20}, using by default  20 re-trainings for the original model and for each of the applied mutations.

The mutation operators in \DC have two types of parameters: continuous and non-continuous.  For example, the operator that removes part of the training data has the continuous parameter \textit{percentage}, which decides what portion of inputs should be deleted. Its value varies in the range  0\% to 99\% (as we cannot delete all training data). In contrast, mutation operators that operate on a per-layer basis have a non-continuous parameter \textit{layer}, which determines the \textit{specific layer} of a neural network to mutate.

\begin{table}[t]
\centering
\scriptsize
\caption{Mutation Operators provided by \DC \cite{DeepCrime} and not killed by the initial test sets of our case studies}
\resizebox{\columnwidth}{!}{%
\begin{tabular}{@{}m{1.3cm}|m{3.9cm}|m{3.4cm}}
\hline
\textbf{Group} & \textbf{Mutation Operator} & \textbf{Mutation Parameters} \\
\hline
\multirow{5}{*}{Training Data} & 	Change labels of training data (TCL)			  & \textit{label} to perform the mutation on  \newline 
															\textit{percentage} of data to mutate \\ \cline{2-3}
		 	& 	Remove portion of training data (TRD)			  & \textit{percentage} of data to delete \\ \cline{2-3}
		 	& 	Unbalance training data (TUD)				  & \textit{percentage} of data to remove\\ \cline{2-3}

		 	& 	Add noise to training data (TAN)				  & \textit{percentage} of data to mutate \\ \cline{2-3}
		 	& 	Make output classes overlap (TCO)			  & \textit{percentage} of data to mutate \\ \hline
\multirow{2}{*}{Hyperparams}	 
	 	& 	Decrease learning rate (HLR)				  & \textit{new learning rate} value\\ \cline{2-3}
	 	& 	Change number of epochs (HNE)			  & \textit{new number of training epochs} \\ \cline{2-3}
	 	\hline
\multirow{4}{*}{Activation}	 	& 	Change activation function (ACH)			  & \textit{layer} w/ non-linear activ. function \newline
															 \textit{new activation function} \\ \cline{2-3}

	 	& 	Remove activation function (ARM)			  & \textit{layer} w/ non-linear activ. function \\ \cline{2-3}
	 	& 	Add activation function to layer (AAL)			  & \textit{layer} w/ linear activation function \newline
															\textit{new activation function} \\ \hline

Regularisation	 	& 	Add weights regularisation (RAW)			  & \textit{layer} w/o weights regularisation \newline
															\textit{new weights regulariser}\\ \cline{2-3}

		 \hline
Weights	 		 	& 	Change weights initialisation (WCI)			  & \textit{layer} to perform the mutation on\newline 
															\textit{new weights initialiser} \\ \cline{2-3}
	 		 \hline
Optimisation	& 	Change optimisation function (OCH)			  & \textit{new optimisation function} \\ \cline{2-3}
\hline
\end{tabular}
}
\label{tab:muoperators}
\end{table}

In case the parameter values are not specified by a user, \DC automatically computes the best configuration for the mutation operator. For non-continuous parameters, \DC performs an exhaustive search by iterating through all of the possible values for a parameter. In the case of continuous parameters, the computation is based on identifying the lowest and the highest possible values and performing a binary search in this range. The aim of the search is to discover the most challenging and yet killable configuration of the mutation operator for a given test suite. 
For example, for the operator \textit{remove portion of training data} (\textit{TRD} in Table~\ref{tab:muoperators}) the binary search first checks if the most aggressive configuration (99\%) is killed by the test data. If so, \DC finds the middle point in the range of possible values (49.5\%) and checks it for killability. If the middle point gets killed, the search continues on the lower part of the range (0\% - 49.5\%); otherwise, on the upper half of the range (49.5\% - 99\%). This process is applied in a recursive manner till the point when the size of a new range becomes smaller than or equal to the desired precision $\epsilon$. The observed value of the percentage parameter that is not killed, which is $\epsilon$-close to the least aggressive killable configuration, is the output of the binary search: this non killed mutant is the target of test generation. 

The authors of \DC also propose a definition of mutation score per operator. The definition is based on the assumption that training data is a set of inputs to which a trained model is the most sensitive. Given a test set \textit{TS}, its mutation score (\textit{MS}) is the proportion of configurations killed by both test and train set over those killed by the train set. It is calculated as: 


\begin{equation} \label{eq:MS}
MS(MO, TS)  = \frac{|K(MO, TS) \cap K(MO, TRS))|}{|K(MO, TRS)|}
\end{equation}

For example, if for the mutation operator \textit{TRD} the least aggressive killed configuration found by binary search is 10\% for the training data and 25\% for the test data, the mutation score will be computed as: $MS = | [0.25:0.99] | / | [0.10:0.99] |$ = 0.74 / 0.89 =  0.83. The overall mutation score of the test suite is computed as the average of mutation scores across all operators.
The fact that \DC offers a wide selection of mutation operators that are based on real DL faults and that it produces a statistically reliable outcome was the key motivation for us to choose this tool. The list of \DC's mutation operators (with their parameters) that produced mutants that were not killed by the test sets used in our case studies can be found in Table \ref{tab:muoperators} (killed mutants are not the target of \tool's input generation).

\section{The \tool Technique}
\label{technique}

\tool aims to augment an existing test set by extending it with mutant killing inputs that increase its mutation score. The \autoref{algo} describes the main steps implemented in \tool to generate new inputs that kill mutants.

\begin{algorithm}[t]
\begin{small}
\caption{Overall algorithm of \tool}
\label{algo}

\SetKwInOut{Input}{Input}
\SetKwInOut{Output}{Output}
\Input{$ts_o$: original test set\\ $C$: original DL program code \\ $g_{max}$: max number of generations \\ $popsize$: population size \\
$mutop$: mutation operator \\ $n$: number of re-training runs \\ $o$: number of original model instances \\ $m$: number of mutant instances}
\Output{$ts_a$: augmented test set}

\tcc{generate original and mutant instances using \DC} \\

\textit{original model instances M$_o$} $\gets$ $\emptyset$\;
\textit{mutant model instances M$_m$} $\gets$ $\emptyset$\;
\textit{M$_o$, M$_m$} $\gets$ \textsc{\DC} (\textit{C}, $mutop$, $n$, $m$, $o$)\;

\tcc{start evolutionary search}\\

\textit{generation g} $\gets$ 0\;

\textit{archive A} $\gets$ $\emptyset$ \hfill \;

\textit{initial population $P_0$} $\gets$ \textsc{InitPopulation}($M_o$, $popsize$)\;
\textit{population $P$} $\gets$ \textit{$P_0$}\;
\textsc{Evaluate}(\textit{P}, $M_o$, $M_m$)\;
\textit{A} $\gets$ \textsc{UpdateArchive}(\textit{P})\;
\tcc{assign  crowding distance to  individuals}\\
\textit{P} $\gets$ \textsc{Select}($P$, $popsize$)\;

\While{$g$ $<$ $g_{max}$ } {
$g$ $\gets$ $g + 1$\;
\tcc{selection based on dominance/crowding distance}\\
\textit{offspring Q} $\gets$ \textsc{SelTourDCD}($P$, $popsize$) \;
\tcc{substitute most dominated/misbehaving on M$_o$}\\
\textit{P} $\gets$ \textsc{Repopulation}($P$, $P_0$, $A$)\;
\ForEach{$q$ $\in$ Q}{
     $q$ $\gets$ \textsc{Mutate}($q$) \;
}
\textsc{Evaluate}($P \cup Q$, $M_o$, $M_m$)\;
\textit{A} $\gets$ \textsc{UpdateArchive}($P \cup Q$)\;
\textit{P} $\gets$ \textsc{Select}($P \cup Q$, $popsize$)\;
}

\tcc{augment the test set with the archived inputs}\\
\textit{ts} $\gets$ $ts_o \cup A$\;
\Return(\textit{ts})
\end{small}
\end{algorithm}

Starting from the original code of a DL model and an existing test set, \tool leverages \DC to obtain the configurations for which the considered mutation operator is not killed by the original test set (for continuous operators, this is the most aggressive non killed configuration found by binary search).
\DC injects the corresponding mutation into the model's code and produces multiple original model and mutant instances by executing $n$ times the training process on the original and the mutated model's code, respectively (line 4). \tool uses evolutionary search to generate new test inputs that can discriminate the original model instances from the mutated ones. The algorithm is based on NSGA-II~\cite{DebAM02}, a multi-objective evolutionary search algorithm largely used in search-based software testing research~\cite{PanichellaKT18,YooH07,YooH10,MaoHJ16,LakhotiaHM07,RiccioFSE20}. 

After initialising variables $g, A, P_0, P$ (lines 6-13), the evolutionary steps are repeated for a given number of iterations, $g_{max}$. In each iteration, a population of individuals, i.e. test inputs, is evolved, and their behaviour is evaluated against the original and mutant models. The result of such evaluation (lines 10 and 23) is the assignment of fitness values to individuals. Based on fitness values, the best individuals are identified and sorted by means of \textit{crowding distance sorting}~\cite{DebAM02}, a technique that accounts for both dominance between individuals according to the fitness values as well as the distance between individuals that belong to the same dominance front (lines 13 and 25). Then, we use tournament selection to select the surviving individuals $Q$ (line 17), which are mutated by genetic operators (line 21).  
The worst individuals are replaced by means of the repopulation operator, which re-introduces some of the initial seeds ($P_0$) into the current population $P$ (line 19).
When mutation killing inputs are generated, they are stored in an archive (lines 11 and 24). Finally, the archived solutions are used to augment the initial test suite (line 28). The test set improvement can be assessed by re-running \DC to check if the previously non-killed configuration is now killed. 

\tool's evolutionary algorithm rewards individuals that behave correctly on original models and misbehave on mutants. \tool is hybridised with novelty search, as it also rewards individuals that exhibit the diversity of behaviours~\cite{LehmanS11, MarculescuFT2016}.
It uses an archive to store the best solutions found during the search in order to avoid cycling.
It also uses repopulation to escape the stagnation in local optima with the high basin of attraction. Preliminary experiments supported the adoption of our newly proposed non-standard twists in NSGA-II (such as the repopulation operator or the hybridisation with novelty search), as they provide more diverse solutions than the standard algorithm.

\subsection{Model-Based Input Representation}

\tool belongs to the family of model-based test input generators~\cite{Utting12}, i.e., tools  that manipulate a model of the input instead of directly modifying the input data (e.g., pixels). In the following, we will refer to the model used for manipulating inputs as \textit{\genmodel} in order to distinguish it from the models used in the DL training and prediction process.

Input data derived from an {\genmodel} are more likely to be realistic and belong to the input validity domain than data subjected to low-level manipulation~\cite{RiccioFSE20, ZohdinasabISSTA2021}. This implies that \tool is applicable to problems for which an {\genmodel} is available. The development of {\genmodel}s is the standard practice in several domains, such as cyber-physical systems, including safety-critical ones, e.g., automotive~\cite{Larman1997}. Below, we present {\genmodel}s for domains we considered in our experimental evaluation: a vector image format for handwritten digit classifiers and a 3D human eye region model for eye gaze predictors.

\begin{figure}
  \centerline{\includegraphics[width=0.7\linewidth]{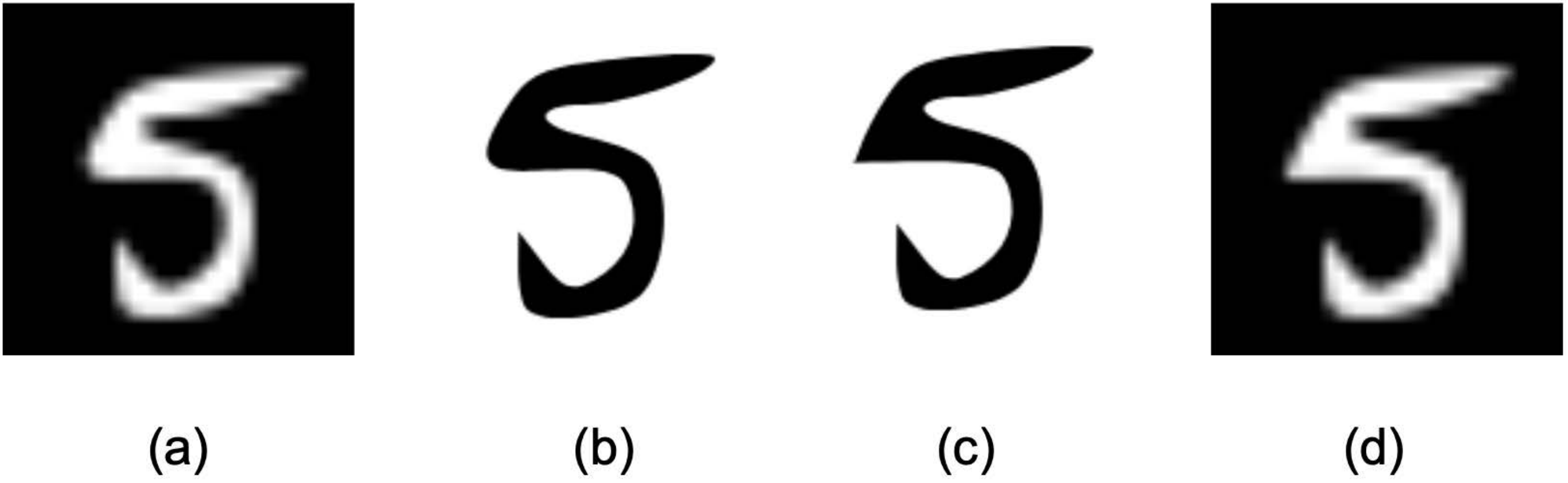}}
  \caption{Digit input representation and mutation. (a)~original input; (b) original SVG model after vectorization; (c) SVG model mutated by moving a control point; (d) mutated input}
  \label{fig:mnist}
\end{figure}

\textit{Digit Classification.} We consider handwritten digit samples in the format adopted by the MNIST database~\cite{LecunBBH98}. Its inputs are originally encoded as $28\times28$ images, with greyscale levels that range from 0 to 255. As shown in~\autoref{fig:mnist}, we model them by adopting Scalable Vector Graphics (SVG)\footnote{\url{https://www.w3.org/Graphics/SVG/}} as their representation. SVG is an XML-based vector image format for two-dimensional graphics, which defines shapes as combinations of cubic and quadratic B\'ezier curves. The control parameters determining the shape of a modelled digit are the start point, the end point and the control points of each B\'ezier curve. This representation helps in preserving the smoothness and curvature of handwritten shapes after minor manipulations of the curve parameters~\cite{RiccioFSE20, ZohdinasabISSTA2021}. We use the Potrace algorithm~\cite{Selinger03} to transform an MNIST input into its SVG model representation. This algorithm performs a sequence of operations to obtain a smooth vector image starting from a bitmap. To transform an SVG model back into a $28\times28$ grayscale image, we perform rasterisation by using two popular open-source libraries (LibRsvg\footnote{\url{https://wiki.gnome.org/Projects/LibRsvg}} and Cairo\footnote{\url{https://www.cairographics.org}}).

\textit{Gaze Prediction.} We focus on the input format for the gaze estimator model proposed by Zhang et al.~\cite{zhang:2015}, which takes as an input an eye image and a \textit{2D} head rotation angle (\textit{pitch} and \textit{yaw}) and predicts the eye gaze angle. The eye images are generated by exploiting \textit{UnityEyes}, a freely available rendering framework~\cite{wood:2016}. Our eye model consists of all the independent parameters used by UnityEyes to generate an eye image. They can be divided into two groups: those that cover various aspects related to an eye appearance  (head angle, eye angle, pupil size, iris size, iris texture, skin texture) and others that describe the lighting (texture, rotation, ambient intensity, exposure for image-based lighting, and rotation and intensity for directional lighting). Only some of these parameters are directly controllable when asking UnityEyes to generate new images, namely head rotation angles and eye rotation angles (the latter providing us with the ground-truth for the gaze prediction), while the others are decided internally by UnityEyes. All parameters are recorded by UnityEyes in a JSON file that accompanies a generated image. For each pair of head and eye angles (controllable parameters), it is possible to request UnityEyes to generate an arbitrary number of eye images, differing among each other by the remaining, not directly controllable, parameters.
When manipulating UnityEyes' parameters for the purpose of test generation, we need to know the range in which each parameter falls in order to ensure the validity of the manipulated values. 
Thus, for head and eye angles, we use the ranges suggested in the \textit{UnityEyes} interface. To learn the valid ranges for the remaining parameters, we generated a dataset of more than 1 million images and analysed the generated \textit{JSON} files. The identified ranges and the script used for such analysis are available in our replication package~\cite{replicationPkg}.

\subsection{Fitness Functions}

\tool optimises two fitness functions, which measure the ability of an individual to kill mutants and its diversity from the solutions already encountered during the search.

\textbf{Mutation Killing.}
The fitness function $f_1$ measures how close an individual is to misbehave on mutants. In particular, for a given mutant instance $mut$, its value is negative in the presence of a misbehaviour, while its value is positive and indicates the distance from a misbehaviour when the system behaves correctly.
We estimate the distance from a misbehaviour as the model's confidence in the predicted class for classifiers or the difference between the tolerable error and the actual prediction error for regressors. Hence, the lower the value assumed by such distance to misbehaviour, the higher the mutant's likelihood of misbehaving. 
To take into account the non-determinism of mutation and training, we generate and train $m$ mutant instances. Correspondingly, the fitness value of an individual is computed as the sum of our misbehaviour closeness metric over $m$ mutant instances. The fitness function $f_1$ has to be minimised:

\begin{equation}
\label{eq:closeness}
\min \; f_1(x) = \min \sum \nolimits_{mut \in M_m}\text{eval$_{mut}$}(x)
\end{equation}
\noindent To compute $f_1$ for an individual $x$, \tool executes the $m$ instances of the considered mutant with $x$ as an input. 
The definition of function \textit{eval} is clearly problem specific.


\textit{Digit Classification.} The \textit{eval} function exploits the classifier's output softmax layer, which can be interpreted as the confidence level assigned to each possible class. The predicted class corresponds to the highest confidence level, and there is a misbehaviour when the expected class has a confidence level lower than another class. In particular, \textit{eval} is calculated as the difference between the confidence associated with the expected class and the maximum confidence associated with any other class when the prediction is correct; it is -1 otherwise.

\textit{Gaze Prediction.} A misbehaviour is detected when the prediction error exceeds the maximum tolerated error. The prediction error is the difference between the model prediction and the expected prediction (provided as ground-truth by UnityEyes). Since predictions consist of a pair of eye rotation angles in radians (pitch and yaw), the error is calculated as the angle between the expected vector and the predicted one. The maximum tolerated error can be set according to problem-specific requirements. In our study, we set it to $5$ degrees, as this is an acceptable error in other gaze prediction applications ~\cite{zhang:2015, DeepCrime}. The value of $f_1$ is the difference between such an acceptable threshold and the actual gaze prediction error.

\textbf{Diversity.}
The fitness function $f_2$ represents an individual's sparseness with respect to individuals in the archive, and we want to maximise it: 
\begin{equation}
\label{eq:ff2}
\max \; f_2(x) = \max \text{spars}(x, A)
\end{equation}
\noindent
where $A$ is the archive of solutions and $x$ is the individual being evaluated. Function \textit{spars} measures the minimum distance of an individual $x$ from the solutions in the archive $A$: $\min_{y \in A, y \neq x}\text{dist}(x, y)$.  
The distance function \textit(dist) is computed on pairs of inputs and is domain-specific. For \textit{Digit Classification}, it is computed as the Euclidean distance between pixel vectors. In the \textit{Gaze Prediction} problem, we use the genotypic distance, i.e., the distance between the chromosomes of two individuals, whose genes are the eye parameters used by UnityEyes. Because in the chromosome there are float, vectorial and categorical gene values, to obtain an overall distance between chromosomes, we compute the distances between genes of the same type, normalise them separately and return the weighted sum of gene distances. In particular: for float genes, we compute the difference $d$ and normalise it as $d / (d+1)$; for the pitch and yaw angles' pairs, we calculate the angle between two vectors in radians (given the natural limits for eye rotation, the difference never exceeds 1 radian); for categorical genes, we assign 0 to the distance if the genes contain the same category, 1 otherwise. 

\subsection{Initial Population}

To obtain the initial population, we first gather a set of seeds, i.e., inputs on which the original models behave correctly. Then, we select the most diverse seeds by computing pairwise distances and greedily constructing the set of most diverse seeds, starting from a randomly selected first seed up to the desired population size. Then, initial individuals are obtained by applying a mutation genetic operator to each selected seed. We considered as seeds the samples in the training set on which the models behave correctly.

\subsection{Archive of Solutions}

The best individuals encountered during the search are kept in the archive of solutions~\cite{DeJong04}. This prevents the search for novelty from \textit{cycling}, a phenomenon where the population moves from one area of the solution space to another and back again, without memory of the areas it has already explored~\cite{Mouret15}. At the end of the last iteration, the archive will contain the final solutions.

An individual of the population is a solution candidate to be included in the archive if it behaves correctly on at least one of the $o$ original model instances and it triggers a misbehaviour on at least one mutant model instance. When a new candidate solution is found, it competes locally with similar solutions already in the archive so that only the best ones are kept, i.e., those with the lowest value of fitness function $f_1$.
 
In the archive used for \textit{Digit Classification}, a solution competes with the archived inputs that are generated from the same MNIST seed. In the archive used for \textit{Gaze Prediction}, we do not rely on the starting seeds, as UnityEyes generates valid eye images from any random vector of controllable parameters within the validity range without requiring to evolve them from an initially valid seed solution. Hence, we had to define a similarity criterion for the archive used for \textit{Gaze Prediction}: if the distance from the nearest neighbour in the archive is higher than a threshold $t_a$, the new individual is kept in the archive. Otherwise, the new candidate competes locally with its nearest neighbour in the archive. 
The threshold $t_a$ is a parameter that can be adjusted by a tester to obtain a proper trade-off between the number of solutions that enter the archive and the diversity of the archive.
To empirically choose the value of $t_a$, we recommend to (1) compute the minimum distance among a randomly selected set of diverse inputs; (2) choose a value greater than this number; (3) iteratively adjust this value based on the corresponding archive size and similarity. 

\begin{figure}
  \centerline{\includegraphics[width=0.8\linewidth]{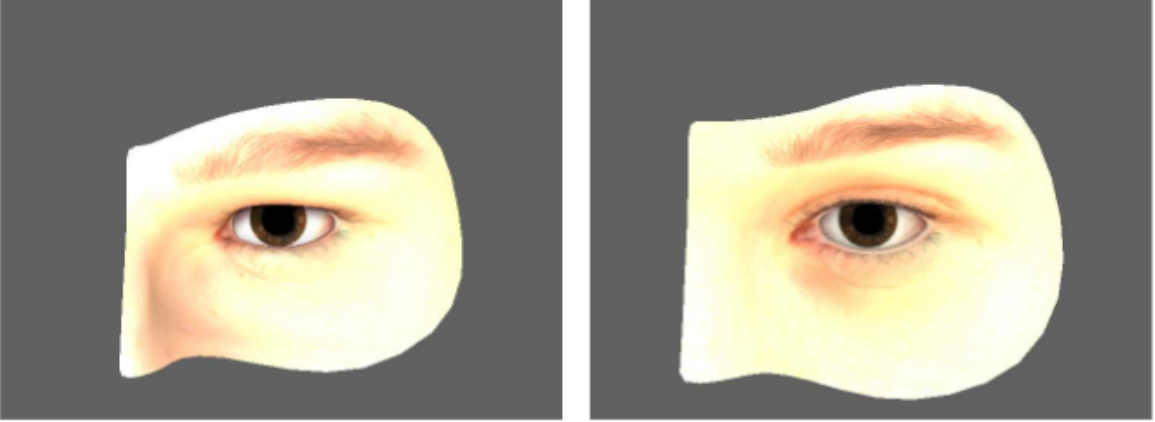}}
  \caption{Eye input mutation}
  \label{fig:eyes}
\end{figure}

\subsection{Genetic Operators}

In multi-objective evolutionary algorithms, there are multiple dimensions (in our case, $f_1$ and $f_2$) on which to compare the individuals. We use the \textsc{Selection} operator from \textsc{NSGA-II}~\cite{DebAM02}, which applies Pareto front analysis and promotes individuals that are not dominated by any other individual. This operator favours individuals with smaller non-domination rank and, when the rank is equal (i.e., they belong to the same Pareto front), it encourages diversity by favouring the one in a less dense region. The offspring of the current population is obtained through tournament selection with the tournament size equal to $2$, by choosing the best between each pair of individuals being compared. 

Each offspring individual is mutated by the \textsc{Mutation} genetic operator, which is domain specific. 
For \textit{Digit Classification}, the mutation genetic operator randomly chooses an SVG model's point and applies a displacement to it in one of the four directions in the 2D space. Then, the rasterisation operation is applied to obtain the new digit image. For \textit{Gaze Prediction}, the mutation genetic operator randomly chooses a gene from the individual's chromosome and applies a displacement to its value. Then, an input image that corresponds to the new values of the eye model's parameters is supposed to be generated. However, since only a small subset of parameters can be controlled in UnityEyes, \tool generates a high number of images and JSON file pairs ($\sim$200) under the desired controllable parameters. From these pairs, it selects the one that is closest to the desired mutant chromosome, checking that it has never been used before during the search. \autoref{fig:eyes} shows an original eye image (left) and the corresponding mutated image (right) obtained by maintaining the controllable parameters unchanged.

During the search, exploration could get stuck in the local optima, despite the use of fitness function $f_2$ to promote diversity.
To mitigate this situation and further vary the population, \tool uses the \textsc{Repopulation} genetic operator,  
which replaces at each iteration the individuals in the population that are behaving incorrectly on all the considered original DL model's instances. The repopulation operator also replaces a fraction of the most dominated individuals in the current population, i.e., the individuals at the bottom of the Pareto front ranking. The aggressiveness of this operator can be tuned by setting the range from which such fraction is uniformly sampled, i.e., the repopulation upper bound. As an example, if the repopulation upper bound is set to $10$, at each iteration, a number \textit{r} is uniformly sampled between 1 and 10, and then the \textit{r} most dominated individuals are replaced.  The new individuals are generated starting from a randomly chosen seed. Repopulation is applied when the archive is not empty.

\section{Experimental Evaluation}
\label{experiment}

\subsection{Subject Systems}

We ran our experiments on two subject systems for which a model of the input is available and can be manipulated via our genetic operators: MNIST and UnityEyes.

MNIST is a publicly available dataset consisting of 70,000 images of hand-written digits. Typically, 60,000 images are used for training and the remaining 10,000 for testing. The DL system consists of a DNN model that predicts which digit is represented by an input image. We considered the deep convolutional neural network (CNN) provided by Keras,\footnote{\url{https://keras.io/examples/vision/mnist_convnet/}} because of its popularity, simplicity and effectiveness ($99.15\%$ test accuracy). 

For the gaze prediction case study based on UnityEyes, we use a multimodal CNN~\cite{unityEyesModel}, which provides an implementation based on the LeNet network architecture~\cite{LecunBBH98} following the approach described in the work by Zhang et al. ~\cite{zhang:2015}. The CNN learns the mapping from an eye image and a \textit{2D} head angle (\textit{pitch} and \textit{yaw}) to a \textit{2D} eye gaze angle. The dataset that we used for training and testing is supplied along with the model and consists of 129,285 eye region images (with 103,428 images used for training and 25,857 for testing) synthesised with \textit{UnityEyes}~\cite{wood:2016}. Each image generated by \textit{UnityEyes} is accompanied by a \textit{JSON} file describing \textit{2D} head angle, eye gaze vector, as well as other parameters used to generate the image, such as skin texture and various lighting features. When presented to a model for training and prediction, the images are converted to grayscale and cropped to $60\times36$ pixels. The head angle and eye angle, which represent the second input to the model and the ground truth, respectively, are converted into radians. 

\subsection{Research Questions}

We have performed a set of experiments to answer the following research questions:

\textbf{RQ1 (Effectiveness):} \textit{Can \tool generate inputs that improve a given test set in terms of mutation killing capability?}

To answer this research question, for each of our subject systems, we need an initial test set that we will then improve with the help of \tool. The original test sets available for these subjects are very large in size and successful in terms of mutation score (100\% for MNIST and 92.5\% for UnityEyes). We, therefore, had to artificially construct a weaker test set for our case studies. For MNIST, we did so by removing the test inputs that are predicted with low confidence (i.e., confidence less than 1) from the original test set. The elimination of such inputs leads to a test set with smaller discriminative power, as low confidence inputs typically represent difficult, corner cases that are effective at discriminating a mutant from the original model. For UnityEyes, which solves a regression, not a classification problem, we instead removed inputs with the smallest standard deviation of loss measured across 20 instances of the original model. Such inputs are very discriminative, as mutants typically amplify the standard deviation of the error observed for the original model, so the effect is more visible when we start from a small standard deviation. A similar approach to construct weak test sets for both classification and regression systems was adopted in Humbatova et al. \cite{DeepCrime}. The approach we used for classification systems has also been previously used in the work by Jahangirova and Tonella~\cite{JahangirovaICST20} for weak test set construction and by Byun et al.~\cite{Byun2019} for test input prioritisation. The size of the weak test set for MNIST is 4,813 elements, and for UnityEyes, it is 4,000 elements.

We then performed mutation testing of our subject systems considering the constructed weak test sets and using \DC. Out of the 24 mutation operators implemented in \DC, 18 were applicable to MNIST  and 17 to UnityEyes. For operators with non-continuous parameters, we applied every value from the list exhaustively. For operators with continuous parameters, we performed the binary search on the full range of the parameter value space. We adopted the statistical notion of mutation killing~\cite{JahangirovaICST20}, using the Wilcoxon test to calculate the $p$-value and Cohen's $d$ to measure the effect size. According to our procedure,  statistical significance is reached when \textit{p\_value} $< 0.05$ and the effect size is greater than ``small'', i.e., Cohen's $d$ $\geq 0.5$. Overall, we got 71 not killed mutants (i.e., mutated versions produced by \DC's mutation operators) for MNIST and 38 for UnityEyes. 

As mutation testing suffers from the problem of equivalent mutants, it is possible that some of the mutants not killed by our weak test set are not killable by any set of inputs, and therefore our attempts for generating inputs that kill these mutants are vain. To avoid this situation, we use the definition of ``likely equivalent'' mutants proposed by Humbatova et al.~\cite{DeepCrime}. According to this definition, if a mutant is not killed by the training data (i.e. the data the mutant should be most sensitive to, as the mutant was trained on such data), then this mutant is deemed \textit{likely equivalent}. After filtering out the likely equivalent mutants, we were left with 19 mutants for MNIST and 10 for UnityEyes. The 19 MNIST mutants belong to 12 different mutation operators, while for UnityEyes 10 mutants are produced by 9 mutation operators. To make our experiments feasible, we further reduced the set of MNIST mutants by picking only one mutant for each mutation operator.

We applied \tool to each of the 22 mutants. We first ran the initial population generation process 10 times to obtain 10 different populations for each subject study. We then invoked the input generation process for each pair of the mutant and initial population, getting as a result 10 runs of \tool on each mutant to account for the non-deterministic search-based nature of our tool. In these experiments, \tool is run in the 1vs5 (1 original vs 5 mutant instances) configuration. 

This means that the number of mutant instances used by the fitness function $f_1$ (see~\autoref{eq:closeness}) is 5. 

The next research question investigates other alternative configurations of our tool. 

\begin{table}
\setlength{\tabcolsep}{8pt}
\renewcommand{\arraystretch}{1.1}
\centering
\caption{\tool Configurations}

\begin{small}
\begin{tabular}{ l r r}

\toprule

 Parameter & MNIST & UnityEyes \\ 
 
 \midrule
 
 population size & 100 & 12 \\  
 generations & 1000 & 100 \\
 archive threshold $t_a$ & - & 0.55 \\
 repopulation upper bound & 10 & 2 \\
 
 \bottomrule
 
\end{tabular}
\end{small}
\label{tab:config}
\end{table}
\textbf{RQ2 (Fitness Guidance)}: \textit{How does the fitness function based on a single mutant instance compare to the fitness function based on multiple mutant instances in guiding \tool towards the generation of mutation killing inputs?}

The aim of this research question is to identify whether providing more instances of the same mutant to \tool increases its success in generating mutation-killing inputs. For this purpose, we ran \tool in 4 different modes by providing it with either 1, 5, 10 or 20 instances of the same mutation (i.e., we configure it as 1vs1, 1vs5, 1vs10 and 1vs20). 

Similarly to RQ1, we perform 10 runs using 10 different initial populations. We do not evaluate extensively the effect of increasing the number of instances of the original model (e.g., 5vs5 or 10vs10), as preliminary experiments showed that the effect of such alternative choices is negligible on the effectiveness of the fitness function, while at the same time is substantially increasing the overall computation time.

\textbf{RQ3 (Comparison with other Tools):} \textit{Can we use existing DL input generators to achieve comparable improvement in the mutation killing capability of a test set?}

To answer this research question, we compare \tool to two state of the art test input generators for DL systems: \textit{DeepJanus} \cite{RiccioFSE20} and \textit{DLFuzz} \cite{GuoJZCS18}. \textit{DeepJanus} is a model-based tool that uses a multi-objective evolutionary algorithm to generate frontier inputs for DL systems. The \textit{frontier inputs} are defined as pairs of inputs that are similar to each other but trigger different behaviours of a DL system. The idea is that for a low-quality DL system, such a frontier will include pairs that intersect the validity domain, while for a high-quality one, it will have a small or no intersection at all. In our experiments, we passed  \textit{DeepJanus} one instance of the original model, and from the generated set of pairs of inputs, we use only those inputs that do not trigger any misbehaviour in the original model, as our goal is to obtain inputs that behave correctly on the original models but misbehave on the mutated ones. Another option could be passing  \textit{DeepJanus} the mutated model and then using the misbehaving set of inputs. However, some preliminary runs showed that the misbehaving inputs for the mutant almost never behave correctly on the original model. Therefore, we excluded this setup from our comparison study. \textit{DeepJanus} can be applied to both UnityEyes and MNIST. Moreover, it shares with \tool the same input representation and mutation genetic operator, which guarantees a fair comparison of the approaches.

\textit{DLFuzz} is representative of search-based fuzzing testing tools that generate test inputs by applying perturbations to the raw input (i.e., pixels)~\cite{Dola21}. It aims to generate adversarial inputs that maximise neuron coverage for a DL system under test. 
For this purpose, \textit{DLFuzz} iteratively selects neurons, the activation of which would lead to increased neuron coverage, and applies perturbations to test inputs in order to activate those neurons, so guiding DL systems towards exposing misbehaviours. The publicly available version of \textit{DLFuzz}\footnote{\url{https://github.com/turned2670/DLFuzz}} does not support regression systems. Therefore we could not apply it to UnityEyes. Moreover, this implementation does not work with Python versions higher than 2.7.1, so we had to update the code to make it compatible with Python 3.8.

Similarly to \tool, both \textit{DeepJanus} and \textit{DLFuzz} are affected by randomness, so we performed 10 runs of each tool, each run using a different initial population. However, we fixed the same population across runs of different tools to ensure that the differences in their performance are not due to the different starting points of the 
algorithms. As explained before, \textit{DeepJanus} uses the original model in its generation process, not requiring a re-run for each mutant. In contrast, as \textit{DLFuzz} generates only inputs that get misclassified by the given DL model, we used the mutants. 
As a result, \textit{DLFuzz} had to be re-run for each considered mutant. 

Overall, we performed 20 runs of \textit{DeepJanus} (10 populations for the original model of both MNIST and UnityEyes), 120 runs of \textit{DLFuzz} (10 populations for 12 MNIST mutants) and 220 runs of \tool (10 populations for 22 MNIST and UnityEyes mutants). 

For both tools, we used the configuration reported as the one achieving the best performance by their authors.




\textbf{RQ4 (Fault Detection):} \textit{Can the test set augmented by DeepMetis expose more faults than the original test set?}

This research question analyses whether \tool delivers its promise of improving the test set so that it detects more faults. Since, to the best of our knowledge, there is no publicly available dataset of reproducible real faults for DL systems, we use \DC mutants as a replacement for real faults in a cross-validation setup. 

Specifically, we perform cross-validation by leaving one of the mutants out and augmenting the test set with all the inputs generated by \tool for the remaining mutants. We ensure that none of the remaining mutants is generated by the same mutation operator as the cross-validation mutant, assuming that mutants produced by the same operator may have similar properties. We then check if the augmented test set is able to kill the cross-validation mutant. This process is repeated separately for the inputs generated in each of the 10 runs of \tool. We added the previously excluded 7 MNIST mutants to this analysis, as,  although there are no inputs generated specifically for them, they can still serve as cross-validation mutants. Before proceeding with the experiment, we performed a redundancy analysis \cite{DeepCrime} among the mutants of each subject to ensure that inputs generated for one mutant do not kill another mutant just because the latter is redundant with respect to the former. Redundancy analysis showed that all 10 UnityEyes mutants are non-redundant, while for MNIST, 6 out of 19 mutants are redundant. We excluded redundant mutants from further analysis, i.e., we did not use them as cross-validation mutants.


\begin{table*}[htb]
\caption{Results: column $K$ (killing probability) reports  mutation score, for continuous operators, and binary killed/non-killed outcome, for discrete operators, both averaged across 10 runs}
\begin{scriptsize}
\begin{center}
\begin{tabular}{c|l|c|cc|cc|cc|cc|cc|cc} \hline
\multirow{3}{*}{\textbf{Subject}}
 & 
 & \textbf{Weak TS}
 & \multicolumn{2}{c|}{\textbf{\tool}} 
 & \multicolumn{2}{c|}{\textbf{\tool}}
 & \multicolumn{2}{c|}{\textbf{\tool}}
 & \multicolumn{2}{c|}{\textbf{\tool}}
 & \multicolumn{2}{c|}{\textbf{DeepJanus}}
 & \multicolumn{2}{c}{\textbf{DLFuzz}}\\
 
 & \multicolumn{1}{c|}{\textbf{MO}}
 & 
 & \multicolumn{2}{c|}{\textbf{(1vs1})} 
 & \multicolumn{2}{c|}{\textbf{(1vs5)}}
 & \multicolumn{2}{c|}{\textbf{(1vs10)}}
 & \multicolumn{2}{c|}{\textbf{(1vs20)}}
 & \multicolumn{2}{c|}{\textbf{}}
 & \multicolumn{2}{c}{\textbf{}}\\
 
& & K &  Inputs  & K &  Inputs  & K &  Inputs  & K &  Inputs  & K &  Inputs  & K  &  Inputs  & K 
\\
\hline
\multirow{13}{*}{MNIST} &
TCL (84.38\%) & 13\% & 21 & 92\% & 16 & 87\% & 18 & 90\% & 20 & 86\% & 8 & 62\% & 61 & 91\%\\
&TRD (89.72\%) & 6\% & 48 & 82\% & 40 & 89\% & 45 & 88\% & 18 & 78\% & 8 & 60\% & 119 & 85\%\\
&TUD (90.62\%) & 6\% & 23 & 78\% & 17 & 77\% & 20 & 73\% & 22 & 73\% & 8 & 18\% & 61 & 68\%\\
&TAN (100\%) & 0\% & 19 & 63\% & 19 & 81\% & 21 & 74\% & 22 & 79\% & 8 & 37\% & 67 & 43\%\\
&TCO (96.88\%) & 0\% & 14 & 49\% & 14 & 59\% & 17 & 69\% & 50 & 60\% & 8 & 29\% & 39 & 48\%\\
&HLR (0.064) & 0\% & 42 & 85\% & 26 & 86\% & 27 & 86\% & 30 & 86\% & 8 & 70\% & 110 & 86\%\\
&HNE (1) & 0\% & 47 & 87\% & 30 & 89\% & 35 & 90\% & 40 & 90\% & 8 & 64\% & 110 & 96\%\\
&ACH (l6; 'sigmoid') & 0\% & 20 & 100\% & 18 & 100\% & 23 & 100\% & 27 & 100\% & 8 & 100\% & 136 & 100\%\\
&ARM (l5) & 0\% & 12 & 90\% & 11 & 100\% & 12 & 100\% & 14 & 90\% & 8 & 10\% & 91 & 100\%\\
&RAW (l0; 'l1\_l2') & 0\% & 15 & 100\% & 11 & 100\% & 15 & 100\% & 16 & 100\% & 8 & 100\% & 52 & 100\%\\
&WCI (l0; 'ones') & 0\% & 21 & 100\% & 14 & 90\% & 24 & 100\% & 28 & 100\% & 8 & 80\% & 170 & 100\%\\
&OCH ('rmsprop') & 0\% & 15 & 100\% & 13 & 100\% & 18 & 100\% & 23 & 100\% & 8 & 100\% & 80 & 100\%\\
& \textbf{Average} & \textbf{2\%} & \textbf{25} & \textbf{86\%} & \textbf{19} & \textbf{89\%} & \textbf{23} & \textbf{89\%} &	\textbf{26} & \textbf{87\%} & \textbf{8} &	\textbf{61\%} & \textbf{91} &	\textbf{85\%}\\
\hline

\multirow{11}{*}{UnityEyes} & TCL (21.88\%) & 86\% & 477 & 86\% & 536 & 88\% & 562 & 86\% & 604 & 86\% & 76 & 86\% & - & - \\
& TRD (41.66\%) & 67\% & 335 & 79\% & 515 & 87\% & 70  & 67\% & 74  & 67\% & 76 & 67\% & - & - \\
& TUD (100\%) & 0\% & 496 & 100\% & 587 & 100\% & 595 & 100\% & 662 & 100\% & 76 & 40\% & - & - \\
& TAN (84.38\%) & 25\% & 379 & 15\% & 546 & 40\% & 669 & 38\% & 611 & 63\% & 76 & 25\% & - & - \\
& HLR (0.0037) & 39\% & 480 & 54\% & 557 & 61\% & 563 & 65\% & 597 & 72\% & 76 & 41\% & - & - \\
& HNE (32) & 70\% & 318 & 70\% & 454 & 72\% & 514 & 69\% & 528 & 70\% & 76 & 70\% & - & - \\
& AAL (l9; 'signsoft') & 0\% & 40  & 0\% & 392  & 90\% & 402  & 60\% & 533 & 60\% &  76 & 0\% & - & - \\
& RAW (l1; 'l2') & 0\% & 48  & 0\% & 480  & 60\% & 517  & 70\% & 557 & 60\% &  76 & 0\% & - & - \\
& RAW (l3; 'l2') & 0\% & 40  & 0\% & 494  & 40\% & 549  & 60\% & 568 & 50\% &  76 & 0\% & - & - \\
& WCI (l1; 'ones') & 0\% & 444  & 0\% & 611  & 40\% & 613  & 60\% & 123 & 0\% &  76 & 0\% & - & - \\
& \textbf{Average} & \textbf{29\%} & \textbf{306} & \textbf{40\%} & \textbf{517} & \textbf{68\%} & \textbf{505} & \textbf{68\%} & \textbf{486} & \textbf{63\%} & \textbf{76} & \textbf{33\%} & \textbf{-} & \textbf{-} \\
\hline
\end{tabular}
\end{center}

\end{scriptsize}
\label{tab:results}
\end{table*}
\subsection{Results}

Columns \textit{Subject} and \textit{MO} in  Table \ref{tab:results} indicate the DL system and the mutation operator that provided the mutants used by \tool for test input generation. For each operator, we report in brackets the parameter values which were found by the binary/exhaustive search and were used to generate the non killed mutant. For mutation operators that manipulate the training data,  this value indicates the ratio of the affected data. For example, \textit{MNIST/TRD} removes 89.72\% of the training data. For the other operators, parameter values with the prefix `l' followed by a number indicate the layer to which a mutation operator was applied. All the other parameters specify the exact value used to inject the fault. For example, \textit{MNIST/ACH (l6; `sigmoid')} means that the activation function of layer number 6 was changed from the original to the `sigmoid' one. 

In  Table \ref{tab:results}, the sub-columns $K$ indicate the \textit{killing probability}, computed as the mutation score (see Equation~(\ref{eq:MS})) for continuous operators or as the binary killed/non-killed outcome for discrete operator (since we did not apply \tool to all the possible mutants produced by discrete operators, Equation~(\ref{eq:MS}) cannot be used for them).
Column \textit{Weak TS} shows the killing probability $K$ of the initial, weak test set. 
In the following columns, the sub-column \textit{Inputs} shows the average number of inputs generated across 10 runs by each tool/tool configuration, while the sub-column \textit{K} shows the average killing probability of the test set augmented with the generated inputs, computed across 10 runs.

\subsubsection{RQ1: Effectiveness}



The results for \tool in its best configuration (1vs5) show that for both subjects, the augmentation of the initial test set with the \tool-generated inputs leads to a substantial increase of the mutation score. For MNIST, the improvement across the operators varies between 59\% and 100\%, with the average $K$ jumping from 2\% to 89\%. For UnityEyes, the improvement ranges from 2\% to 100\% on a per operator basis and the average $K$ rises from 29\% to 68\%. The number of generated inputs, which would require manual labelling, is 19 on average for MNIST and 517 for UnityEyes. As these numbers constitute only 0.0003\% of the training data set size for MNIST and 0.005\% for UnityEyes, we consider the labelling effort associated with \tool to be low.


\begin{tcolorbox}
\textbf{RQ1}: \tool is able to achieve a substantial improvement in killing probability on each of the provided mutants. The magnitude of this improvement is 87\% for MNIST and 39\% for UnityEyes. The manual labelling effort for the newly generated inputs can be deemed acceptable.
\end{tcolorbox}

\subsubsection{RQ2: Fitness Guidance} 

Columns \tool (1vs1), \tool (1vs5), \tool (1vs10), \tool (1vs20) report the results obtained when the fitness function uses 1, 5, 10 and all 20 instances of a mutant during the input generation process, respectively. In the case of MNIST, for 3 mutants out of 12, 1vs1 and 1vs5 provide the same results. For 6 operators, 1vs5 performs better; however, for 2 out of those, the improvement is marginal (1-3\%). For the remaining 3 operators, 1vs1 outperforms 1vs5, with the difference for one of the operators being only 2\%. When we further compare 1vs5 to 1vs10, the latter exhibits an improvement for 4, equal performance for 5 and deterioration for 3 operators while being substantially more expensive computationally. 
Overall, as also reflected in the average $K$ across operators, for MNIST, the optimal performance is obtained with 1vs5 and 1vs10 settings, which provide slightly better results than 1vs1 and 1vs20. 

The results for UnityEyes show that 1vs5 and 1vs10 produce the same average $K$ (68\%), which is slightly better than 1vs20 (63\%), but is definitely superior when compared to 1vs1  (40\%). On a closer inspection, 1vs5 outperforms 1vs10 and 1vs20 on 5 mutants out of 10, with the majority of them being continuous operators, while 1vs10 is the best in 3 cases and 1vs20 in 2. As was noted, 1vs20 on average performs similarly to 1vs5 and 1vs10; however, in one case (\textit{WCI (l1; ’ones’)}), it fails to produce any improvement at all. 

The reason behind the comparative weakness of 1vs1 w.r.t. the other settings is that its fitness function has a very limited range because it aggregates the \textit{eval} value of a single mutant instance, which provides restricted guidance to the test generation process.


The input generation for our experiments was performed on various machines. It complicates the comparison of the execution time between different configurations of \tool. However, for each subject, we ensured to run all 4 configurations on the \textit{TUD} operator (selected randomly) using the same machine. For MNIST, we used a MacBook Pro laptop (2.2 GHz Intel Core i7, 6 cores, 16GB RAM), while for UnityEyes, we used Alienware Aurora R8 (3.60 GHz Intel Core i9-9900K, 8 cores, 32GB RAM, NVIDIA GeForce RTX 2080 Ti 11 GB). For MNIST, this operator took 6, 22, 47 and 57 minutes on average across 10 runs for 1vs1, 1vs5, 1vs10 and 1vs20, respectively. For UnityEyes, the generation of inputs for one run on average lasted 53 (1vs1), 66 (1vs5), 65 (1vs10), and 69 (1vs20) minutes. These results show that 1vs5 is the optimal setting for balancing the improvement in mutation score and the time required to generate the inputs.

\begin{tcolorbox}
\textbf{RQ2}: The 1vs5 configuration of \tool proved to be the optimal one. It outperforms 1vs1 by a substantial margin, as a single mutant instance (1vs1) cannot provide enough guidance to generate effective inputs. The settings with a  higher number of mutant instances are sometimes comparable in terms of mutation score improvement, but they might require significantly more computation time.



\end{tcolorbox}
\subsubsection{RQ3: Comparison with other Tools}

Columns \textit{DLFuzz} and \textit{DeepJanus} in Table \ref{tab:results} report the results for each of the tools being compared to \tool. In the case of MNIST, for 9 out of 12 mutants, \tool (1vs5) performs better than \textit{DeepJanus}, while for the remaining mutants, they have similar performance. The average $K$ across all mutants for \tool (1vs5) is higher by 28\% than for DeepJanus. When it comes to the comparison between \tool and \textit{DLFuzz}, \tool provides better results for 4 mutants, \textit{DLFuzz} for 3 mutants, and the outcome is equal for the remaining 5. The average $K$ across all mutants is 89\% for \tool (1vs5) and 85\% for \textit{DLFuzz}. However, \textit{DLFuzz} generates 4.8 more inputs than \tool (1vs5) and therefore requires much more manual labelling effort.

As \textit{DLFuzz} is not applicable to regression problems, the comparison for the UnityEyes subject was only possible between \tool and \textit{DeepJanus}.  Results show that \textit{DeepJanus} is not able to produce any improvement in the majority of the cases. The only exceptions are \textit{TUD} and \textit{HLR} operators, where for the former, the average improvement is 40\% compared to 100\% of \tool (1vs5), and for the latter, the improvement of \textit{DeepJanus} is limited to 2\% vs 22\% of \tool.

We performed statistical analysis on the comparison of the results by each tool. For mutants with continuous parameters, we used the Wilcoxon statistical test to obtain the $p$-value and the Vargha-Delaney $\hat{A}_{12}$ to quantify the effect size. For mutants with non-continuous parameters, we calculate confidence intervals using Wilson's method. When comparing \tool and \textit{DeepJanus} for MNIST, the difference is statistically significant ($p$-value $<$ 0.05 or confidence intervals do not intersect) for 7 mutants out of 12. For 5 out of 7 mutants with continuous parameters, the effect size is large; for 1 mutant, it is medium, and for the remaining one, it is small. In case of \tool and \textit{DLFuzz}, there is a statistically significant difference for 2 mutants. The effect size is negligible for 1, small for 3, medium for 2 and large for 1 mutant. The results of the comparison of \tool (1vs5) and \textit{DeepJanus} on the UnityEyes subject are statistically significant for 5 out of 10 applied mutants. For the 6 mutants with continuous parameters, the effects size ranges between large (3), small (2), and negligible (1).

When it comes to execution time comparison (conducted in the same conditions as described for RQ2), for MNIST \tool took on average 22 minutes, \textit{DeepJanus} 9 minutes and  DLFuzz 24 minutes. For UnityEyes, \tool took about 66 minutes on average and \textit{DeepJanus} about 86 minutes.

\begin{tcolorbox}
\textbf{RQ3}: \tool outperforms \textit{DLFuzz} and \textit{DeepJanus} in the task of augmenting a test set to improve its mutation score. 
\end{tcolorbox}

\subsubsection{RQ4: Fault Detection}
Results are presented in Table \ref{tab:fault_detection}, where column \textit{MO} specifies the cross-validation mutant, used to check the hypothesis that \tool generated inputs are also able to kill other, previously unseen mutants. 
Column \textit{Inputs} indicates the average number of inputs that were generated by \tool and added to the originally weak test set across 10 runs. Finally, column \textit{Killed} reports the proportion of runs (out of 10) in which the augmented test set was able to kill the validation mutant. 
 
\begin{table}[htb]
\caption{Fault Detection}
\begin{small}
\begin{center}
\begin{tabular}{l|l|cc} \hline
\textbf{Subject} & \textbf{MO} &  \textbf{Inputs}  & \textbf{Killed} \\
\hline
\multirow{13}{*} {MNIST} 
& TCL (84.38\%) & 107 & 10/10 \\
& TUD (90.62\%) & 106 & 10/10 \\
& TCO (96.88\%) & 109 & 10/10 \\
& HLR (0.064) & 98 & 10/10 \\
& ACH (l6; 'hard\_sigmoid') & 123 &  10/10\\
& ACH (l6; 'softplus') & 123 &  10/10\\
& ACH (l6; 'softmax') & 123 &  10/10\\
& ARM (l5) & 112 & 8/10 \\
& RAW (l0; 'l1\_l2') & 112 & 10/10 \\
& RAW (l0; 'l2') & 112 & 8/10 \\
& WCI (l0; 'ones') & 109 & 10/10 \\
& WCI (l0; 'random\_uniform') & 109 & 1/10 \\
& OCH ('rmsprop') & 111 & 10/10 \\
\hline
\multirow{10}{*} {UnityEyes}
& TCL (21.88\%) & 4635 & 10/10 \\
& TRD (46.41\%) & 4655 & 1/10 \\
& TUD (100\%) & 4584 & 10/10 \\
& TAN (84.38\%) & 4625 & 10/10 \\
& HLR (0.0037) & 4613 & 10/10 \\
& HNE (32) & 4717 & 3/10 \\
& AAL (l9; 'signsoft') & 4779 & 6/10 \\
& RAW (l1; 'l2') & 4197 & 10/10 \\
& RAW (l3; 'l2') & 4197 & 5/10 \\
& WCI (l1; 'ones') & 4559 & 6/10 \\
\hline
\end{tabular}
\end{center}
\end{small}
\label{tab:fault_detection}
\end{table}

For MNIST, almost all validation mutants were killed in all 10 runs, with the exception of ARM (l5) and RAW (l0, 'l2') that were killed in 8 runs and WCI (l0; ’random\_uniform’) that was killed in 1 run. 
The latter is an almost equivalent mutant, with a very low triviality score~\cite{DeepCrime}, which is very difficult to kill for \tool.
The results for UnityEyes also indicate that \tool is always able to kill the unseen mutant at least once. For 5 mutants out of 10, the test set augmented with \tool inputs killed the mutant in 100\% of the runs. In all other cases except for \textit{TRD (46.41\%)} and \textit{HNE (32)}, the augmented test set succeeds in 5 to 6 out of 10 runs.

\begin{tcolorbox}
\textbf{RQ4}: The mutation killing capability of the \tool-generated inputs holds also for previously unseen mutants, with 82\% average success rate across our two subjects.
\end{tcolorbox}

\subsection{Threats to Validity}

\textbf{Construct Validity}: The choice of the distance metrics may threaten our findings. We chose sound metrics for the considered domains. We used Euclidean distance when comparing matrices of grayscale values (also used in previous studies~\cite{RiccioFSE20}).  When comparing UnityEyes inputs, we used a combination of appropriate distances for each gene type in the chromosome. 

\textbf{Internal Validity}: 
The main threat affecting the internal validity of our results is the choice of mutation operators and mutation tool.
We use \DC, a DL mutation tool that accounts for the stochastic nature of DL systems and DL specific mutation operators by adopting the statistical notion of mutation killing. Moreover, its operators are derived from real DL faults that ensure a higher degree of realism as compared to alternatives.

\textbf{External Validity}: The choice of the subject DL systems is a possible threat to the \textit{external validity}. To mitigate it, we chose two diverse DL systems. One solves a classification problem, while another solves a regression problem. The execution of multiple original and mutant models may hinder the generalisation to more complex DL problems, e.g. self-driving cars that require simulations to be evaluated. However, our results show that \tool generates effective inputs with a limited number of models, i.e., 1 original and 5 mutated. A wider set of systems (including industrial ones) should be considered in future studies to further generalise our findings. 

To ensure \textbf{Reproducibility} of our results, we share online the source code of \tool, the considered subjects, and the experimental data~\cite{replicationPkg}.

\section{Related Work}
\label{related}
\subsection{Test Generation for DL Systems}

Several works in the literature~\cite{PeiCYJ17,GuoJZCS18, TianPSB18, ZhangZZ0K18, LeeISSTA20} propose techniques that generate test inputs for DL systems by manipulating raw input data, i.e. they apply small perturbations to available real inputs. A limitation of these approaches is the lack of realism of the generated inputs. While these corrupted images are useful for security testing as adversarial attacks, they are not necessarily representative of data captured by sensors of a real DL system. 

Another family of testing techniques~\cite{AbdessalemNBS16, AbdessalemPNBS18, AbdessalemNBS18,GambiMF19,RiccioFSE20,UdeshiAC18, ZohdinasabISSTA2021} adopts a model-based approach that exploits model manipulation and model-based generation. Differently from raw input manipulation approaches, these techniques tend to generate more realistic inputs if a faithful model of the input domain is adopted since the generated images are compliant with the constraints of such a model. In this work, we adopt a model-based approach to improve the realism of the generated inputs.

Pei et al. propose a raw input manipulation technique aimed at generating inputs that trigger inconsistencies between multiple DL systems~\cite{PeiCYJ17}. Other techniques manipulate raw images and consider as failures the inconsistent behaviours triggered by the original and transformed test inputs~\cite{GuoJZCS18, TianPSB18, ZhangZZ0K18, LeeISSTA20}. 

Model-based approaches, proposed by Abdessalem et al.~\cite{AbdessalemNBS16, AbdessalemPNBS18, AbdessalemNBS18} and Gambi et al.~\cite{GambiMF19}, aim to test advanced driver-assistance systems by generating extreme and challenging scenarios that maximise the number of detected system failures. Riccio and Tonella proposed a model-based approach that produces test suites made of pairs of inputs that identify the frontier of behaviours of a DL system, i.e. the inputs at which the DL system starts to misbehave~\cite{RiccioFSE20}. Udeshi et al. generate inputs that highlight fairness violations by perturbing discriminatory parameters, e.g. gender~\cite{UdeshiAC18}. Vahdat Pour et al.~\cite{PourICST21} use DL mutation to guide the generation of adversarial code snippets for DL models tailored to the computation of code embeddings.

\tool differs from the existing approaches because its goal is to increase the mutation killing ability of a test set.
With the advent of DL mutation frameworks such as DeepMutation~\cite{Ma:2018}, MuNN~\cite{Shen:2018} and DeepCrime~\cite{DeepCrime}, the problem of achieving a high mutation score is increasingly important, especially when mutants mimic real faults, as is the case of DeepCrime~\cite{DeepCrime}. 

\tool is the first approach that can assist developers in the challenging task of making a DL test set better at mutation killing.

\subsection{Test Adequacy for DL Systems}

Several test adequacy criteria have been proposed for DL systems. Pei et al.~\cite{PeiCYJ17} use the number of neuron activations of the model to measure test adequacy. In particular, a neuron is considered activated if its output value is higher than a predefined threshold.  Ma et al.~\cite{Ma-ASE-2018} propose a set of additional adequacy criteria based on neuron activations. They use activation values obtained from the training data and divide the range of values for each neuron into $k$ buckets. Kim et al.~\cite{KimFY19} designed a test adequacy criterion, named \textit{surprise adequacy}, based on the degree of ``surprise'' of an input for the neural network. Similarly to Ma et al.'s criteria~\cite{Ma-ASE-2018}, bucketing is used to make the surprise measure an adequacy criterion: all $k$ buckets of surprise ranges must be covered by the test set. X. Zhang et al.~\cite{ZhangICSE20} observe how inputs are distributed across different uncertainty patterns, i.e. combinations of alternative uncertainty metrics (e.g., high prediction confidence and low variation ratio). Although they do not define a proper adequacy criterion, they recommend generating additional test inputs to cover the least covered uncertainty patterns, and they show that such inputs evade defences against adversarial attacks. 

We adopt a test set's mutation score as an adequacy criterion.
Like other criteria~\cite{PeiCYJ17, Ma-ASE-2018, KimFY19}, our criterion uses the training set as a  reference since it contains the inputs to which the model is mostly sensitive: the mutation score of a test set should be as close as possible to the training set's one.

Jahangirova \& Tonella~\cite{JahangirovaICST20} compared mutation score to other adequacy metrics such as neuron coverage~\cite{PeiCYJ17} and surprise coverage~\cite{KimFY19}, showing that mutation score is more effective in differentiating between weak and strong test sets than the existing alternatives. 

\tool is the first tool that uses mutation adequacy as guidance for the generation of inputs that increase the mutation score of an existing weak test set. 

%

\section{Conclusions and Future Work}
\label{conclusions}

We proposed \tool, the first automated test generator for DL systems that can increase the mutation score of a weak test set, guided by mutation adequacy. 
Our empirical evaluation shows that our tool outperforms state-of-the-art DL test generators in this task. The test sets generated by \tool can expose unknown faults, simulated in our leave-one-out experiment by means of previously unseen mutants. In our future work, we plan to
generalise our results to a wider sample of DL systems, including industrial ones.

\section*{Acknowledgment}
This work was partially supported by the H2020 project PRECRIME, funded under the ERC Advanced Grant 2017 Program (ERC Grant Agreement n. 787703).


\balance
\bibliographystyle{IEEEtran}
\bibliography{biblio.bib}

\begin{thebibliography}{10}
\providecommand{\url}[1]{#1}
\csname url@samestyle\endcsname
\providecommand{\newblock}{\relax}
\providecommand{\bibinfo}[2]{#2}
\providecommand{\BIBentrySTDinterwordspacing}{\spaceskip=0pt\relax}
\providecommand{\BIBentryALTinterwordstretchfactor}{4}
\providecommand{\BIBentryALTinterwordspacing}{\spaceskip=\fontdimen2\font plus
\BIBentryALTinterwordstretchfactor\fontdimen3\font minus
  \fontdimen4\font\relax}
\providecommand{\BIBforeignlanguage}[2]{{%
\expandafter\ifx\csname l@#1\endcsname\relax
\typeout{** WARNING: IEEEtran.bst: No hyphenation pattern has been}%
\typeout{** loaded for the language `#1'. Using the pattern for}%
\typeout{** the default language instead.}%
\else
\language=\csname l@#1\endcsname
\fi
#2}}
\providecommand{\BIBdecl}{\relax}
\BIBdecl

\bibitem{Manning-IIR-2008}
C.~D. Manning, P.~Raghavan, and H.~Sch\"{u}tze, \emph{Introduction to
  Information Retrieval}.\hskip 1em plus 0.5em minus 0.4em\relax New York, NY,
  USA: Cambridge University Press, 2008.

\bibitem{PeiCYJ17}
K.~Pei, Y.~Cao, J.~Yang, and S.~Jana, ``Deepxplore: Automated whitebox testing
  of deep learning systems,'' in \emph{Proceedings of the 26th Symposium on
  Operating Systems Principles}.\hskip 1em plus 0.5em minus 0.4em\relax ACM,
  2017, pp. 1--18.

\bibitem{GuoJZCS18}
J.~Guo, Y.~Jiang, Y.~Zhao, Q.~Chen, and J.~Sun, ``Dlfuzz: differential fuzzing
  testing of deep learning systems,'' in \emph{Proceedings of the 2018 {ACM}
  Joint Meeting on European Software Engineering Conference and Symposium on
  the Foundations of Software Engineering, {ESEC/SIGSOFT} {FSE}}, 2018, pp.
  739--743.

\bibitem{TianPSB18}
\BIBentryALTinterwordspacing
Y.~Tian, K.~Pei, S.~Jana, and B.~Ray, ``Deeptest: Automated testing of
  deep-neural-network-driven autonomous cars,'' in \emph{Proceedings of the
  40th International Conference on Software Engineering}, ser. ICSE '18.\hskip
  1em plus 0.5em minus 0.4em\relax New York, NY, USA: ACM, 2018, pp. 303--314.
  [Online]. Available: \url{http://doi.acm.org/10.1145/3180155.3180220}
\BIBentrySTDinterwordspacing

\bibitem{XieISSTA19}
\BIBentryALTinterwordspacing
X.~Xie, L.~Ma, F.~Juefei-Xu, M.~Xue, H.~Chen, Y.~Liu, J.~Zhao, B.~Li, J.~Yin,
  and S.~See, ``Deephunter: A coverage-guided fuzz testing framework for deep
  neural networks,'' in \emph{Proceedings of the 28th ACM SIGSOFT International
  Symposium on Software Testing and Analysis}, ser. ISSTA 2019.\hskip 1em plus
  0.5em minus 0.4em\relax New York, NY, USA: Association for Computing
  Machinery, 2019, p. 146–157. [Online]. Available:
  \url{https://doi.org/10.1145/3293882.3330579}
\BIBentrySTDinterwordspacing

\bibitem{KimFY19}
J.~Kim, R.~Feldt, and S.~Yoo, ``Guiding deep learning system testing using
  surprise adequacy,'' in \emph{Proceedings of the 41st International
  Conference on Software Engineering, {ICSE}}, 2019, pp. 1039--1049.

\bibitem{Harel-CanadaWGG20}
F.~Harel{-}Canada, L.~Wang, M.~A. Gulzar, Q.~Gu, and M.~Kim, ``Is neuron
  coverage a meaningful measure for testing deep neural networks?'' in
  \emph{{ESEC/FSE} '20: 28th {ACM} Joint European Software Engineering
  Conference and Symposium on the Foundations of Software Engineering, Virtual
  Event, USA, November 8-13, 2020}, 2020, pp. 851--862.

\bibitem{JiaTSE11}
Y.~{Jia} and M.~{Harman}, ``An analysis and survey of the development of
  mutation testing,'' \emph{IEEE Transactions on Software Engineering},
  vol.~37, no.~5, pp. 649--678, 2011.

\bibitem{HumbatovaICSE20}
N.~Humbatova, G.~Jahangirova, G.~Bavota, V.~Riccio, A.~Stocco, and P.~Tonella,
  ``Taxonomy of real faults in deep learning systems,'' in \emph{Proceedings of
  42nd International Conference on Software Engineering}, ser. ICSE
  ’20.\hskip 1em plus 0.5em minus 0.4em\relax ACM, 2020, p. 12 pages.

\bibitem{Zhang20}
J.~M. {Zhang}, M.~{Harman}, L.~{Ma}, and Y.~{Liu}, ``Machine learning testing:
  Survey, landscapes and horizons,'' \emph{IEEE Transactions on Software
  Engineering}, vol. Early Access, no.~--, pp. 1--1, 2020.

\bibitem{RiccioEMSE20}
\BIBentryALTinterwordspacing
V.~Riccio, G.~Jahangirova, A.~Stocco, N.~Humbatova, M.~Weiss, and P.~Tonella,
  ``Testing machine learning based systems: a systematic mapping,''
  \emph{Empir. Softw. Eng.}, vol.~25, no.~6, pp. 5193--5254, 2020. [Online].
  Available: \url{https://doi.org/10.1007/s10664-020-09881-0}
\BIBentrySTDinterwordspacing

\bibitem{JahangirovaICST20}
G.~Jahangirova and P.~Tonella, ``An empirical evaluation of mutation operators
  for deep learning systems,'' in \emph{IEEE International Conference on
  Software Testing, Verification and Validation}, ser. ICST'20.\hskip 1em plus
  0.5em minus 0.4em\relax IEEE, 2020, p. 12 pages.

\bibitem{Sohn:2019}
J.~Sohn, S.~Kang, and S.~Yoo, ``Search based repair of deep neural networks,''
  \emph{arXiv preprint arXiv:1912.12463}, 2019.

\bibitem{Wang:2019}
J.~Wang, G.~Dong, J.~Sun, X.~Wang, and P.~Zhang, ``Adversarial sample detection
  for deep neural network through model mutation testing,'' in \emph{2019
  IEEE/ACM 41st International Conference on Software Engineering (ICSE)}.\hskip
  1em plus 0.5em minus 0.4em\relax IEEE, 2019, pp. 1245--1256.

\bibitem{PourICST21}
M.~V. Pour, Z.~Li, L.~Ma, and H.~Hemmati, ``A search-based testing framework
  for deep neural networks of source code embedding,'' in \emph{IEEE
  International Conference on Software Testing, Verification and Validation},
  ser. ICST'21.\hskip 1em plus 0.5em minus 0.4em\relax IEEE, 2021, p. 11 pages.

\bibitem{QualityMetrics2021}
J.~Gunel, S.~Andrea, and T.~Paolo, ``Quality metrics and oracles for autonomous
  vehicles testing,'' in \emph{2021 IEEE 14th International Conference on
  Software Testing, Validation and Verification (ICST)}.\hskip 1em plus 0.5em
  minus 0.4em\relax IEEE, 2021.

\bibitem{Ma:2018}
\BIBentryALTinterwordspacing
L.~Ma, F.~Zhang, J.~Sun, M.~Xue, B.~Li, F.~Juefei{-}Xu, C.~Xie, L.~Li, Y.~Liu,
  J.~Zhao, and Y.~Wang, ``Deepmutation: Mutation testing of deep learning
  systems,'' in \emph{29th {IEEE} International Symposium on Software
  Reliability Engineering, {ISSRE} 2018, Memphis, TN, USA, October 15-18,
  2018}, 2018, pp. 100--111. [Online]. Available:
  \url{https://doi.org/10.1109/ISSRE.2018.00021}
\BIBentrySTDinterwordspacing

\bibitem{Shen:2018}
W.~{Shen}, J.~{Wan}, and Z.~{Chen}, ``Munn: Mutation analysis of neural
  networks,'' in \emph{2018 IEEE International Conference on Software Quality,
  Reliability and Security Companion (QRS-C)}, July 2018, pp. 108--115.

\bibitem{Ma:2019}
Q.~Hu, L.~Ma, X.~Xie, B.~Yu, Y.~Liu, and J.~Zhao, ``Deepmutation++: A mutation
  testing framework for deep learning systems,'' in \emph{2019 34th IEEE/ACM
  International Conference on Automated Software Engineering (ASE)}.\hskip 1em
  plus 0.5em minus 0.4em\relax IEEE, 2019, pp. 1158--1161.

\bibitem{DeepCrime}
N.~Humbatova, G.~Jahangirova, and P.~Tonella, ``Deepcrime: Mutation testing of
  deep learning systems based on real faults,'' in \emph{Proceedings of the
  30th ACM SIGSOFT International Symposium on Software Testing and Analysis},
  2021.

\bibitem{Islam:2019}
\BIBentryALTinterwordspacing
M.~J. Islam, G.~Nguyen, R.~Pan, and H.~Rajan, ``A comprehensive study on deep
  learning bug characteristics,'' in \emph{Proceedings of the 2019 27th ACM
  Joint Meeting on European Software Engineering Conference and Symposium on
  the Foundations of Software Engineering}, ser. ESEC/FSE 2019.\hskip 1em plus
  0.5em minus 0.4em\relax New York, NY, USA: ACM, 2019, pp. 510--520. [Online].
  Available: \url{http://doi.acm.org/10.1145/3338906.3338955}
\BIBentrySTDinterwordspacing

\bibitem{Zhang:2018}
\BIBentryALTinterwordspacing
Y.~Zhang, Y.~Chen, S.-C. Cheung, Y.~Xiong, and L.~Zhang, ``An empirical study
  on tensorflow program bugs,'' in \emph{Proceedings of the 27th ACM SIGSOFT
  International Symposium on Software Testing and Analysis}, ser. ISSTA
  2018.\hskip 1em plus 0.5em minus 0.4em\relax New York, NY, USA: ACM, 2018,
  pp. 129--140. [Online]. Available:
  \url{http://doi.acm.org/10.1145/3213846.3213866}
\BIBentrySTDinterwordspacing

\bibitem{DebAM02}
K.~{Deb}, A.~{Pratap}, S.~{Agarwal}, and T.~{Meyarivan}, ``A fast and elitist
  multiobjective genetic algorithm: Nsga-ii,'' \emph{IEEE Transactions on
  Evolutionary Computation}, vol.~6, no.~2, pp. 182--197, April 2002.

\bibitem{PanichellaKT18}
A.~Panichella, F.~M. Kifetew, and P.~Tonella, ``Automated test case generation
  as a many-objective optimisation problem with dynamic selection of the
  targets,'' \emph{{IEEE} Transactions on Software Engineering}, vol.~44,
  no.~2, pp. 122--158, 2018.

\bibitem{YooH07}
\BIBentryALTinterwordspacing
S.~Yoo and M.~Harman, ``Pareto efficient multi-objective test case selection,''
  in \emph{Proceedings of the 2007 International Symposium on Software Testing
  and Analysis}, ser. ISSTA '07.\hskip 1em plus 0.5em minus 0.4em\relax New
  York, NY, USA: ACM, 2007, pp. 140--150. [Online]. Available:
  \url{http://doi.acm.org/10.1145/1273463.1273483}
\BIBentrySTDinterwordspacing

\bibitem{YooH10}
\BIBentryALTinterwordspacing
------, ``Using hybrid algorithm for pareto efficient multi-objective test
  suite minimisation,'' \emph{Journal of Systems and Software}, vol.~83, no.~4,
  pp. 689 -- 701, 2010. [Online]. Available:
  \url{http://www.sciencedirect.com/science/article/pii/S0164121209003069}
\BIBentrySTDinterwordspacing

\bibitem{MaoHJ16}
\BIBentryALTinterwordspacing
K.~Mao, M.~Harman, and Y.~Jia, ``Sapienz: Multi-objective automated testing for
  android applications,'' in \emph{Proceedings of the 25th International
  Symposium on Software Testing and Analysis}, ser. ISSTA 2016.\hskip 1em plus
  0.5em minus 0.4em\relax New York, NY, USA: ACM, 2016, pp. 94--105. [Online].
  Available: \url{http://doi.acm.org/10.1145/2931037.2931054}
\BIBentrySTDinterwordspacing

\bibitem{LakhotiaHM07}
\BIBentryALTinterwordspacing
K.~Lakhotia, M.~Harman, and P.~McMinn, ``A multi-objective approach to
  search-based test data generation,'' in \emph{Proceedings of the 9th Annual
  Conference on Genetic and Evolutionary Computation}, ser. GECCO '07.\hskip
  1em plus 0.5em minus 0.4em\relax New York, NY, USA: ACM, 2007, pp.
  1098--1105. [Online]. Available:
  \url{http://doi.acm.org/10.1145/1276958.1277175}
\BIBentrySTDinterwordspacing

\bibitem{RiccioFSE20}
\BIBentryALTinterwordspacing
V.~Riccio and P.~Tonella, ``Model-based exploration of the frontier of
  behaviours for deep learning system testing,'' in \emph{Proceedings of the
  28th ACM Joint Meeting on European Software Engineering Conference and
  Symposium on the Foundations of Software Engineering}, ser. ESEC/FSE
  2020.\hskip 1em plus 0.5em minus 0.4em\relax New York, NY, USA: Association
  for Computing Machinery, 2020, p. 876–888. [Online]. Available:
  \url{https://doi.org/10.1145/3368089.3409730}
\BIBentrySTDinterwordspacing

\bibitem{LehmanS11}
\BIBentryALTinterwordspacing
J.~Lehman and K.~O. Stanley, ``Abandoning objectives: Evolution through the
  search for novelty alone,'' \emph{Evolutionary Computation}, vol.~19, no.~2,
  pp. 189--223, 2011. [Online]. Available:
  \url{https://doi.org/10.1162/EVCO_a_00025}
\BIBentrySTDinterwordspacing

\bibitem{MarculescuFT2016}
B.~{Marculescu}, R.~{Feldt}, and R.~{Torkar}, ``Using exploration focused
  techniques to augment search-based software testing: An experimental
  evaluation,'' in \emph{2016 IEEE International Conference on Software
  Testing, Verification and Validation (ICST)}, April 2016, pp. 69--79.

\bibitem{Utting12}
M.~Utting, A.~Pretschner, and B.~Legeard, ``A taxonomy of model-based testing
  approaches,'' \emph{Software testing, verification and reliability}, vol.~22,
  no.~5, pp. 297--312, 2012.

\bibitem{ZohdinasabISSTA2021}
T.~Zohdinasab, V.~Riccio, A.~Gambi, and P.~Tonella, ``Deephyperion: exploring
  the feature space of deep learning-based systems through illumination
  search,'' in \emph{Proceedings of the 30th ACM SIGSOFT International
  Symposium on Software Testing and Analysis}, 2021, pp. 79--90.

\bibitem{Larman1997}
C.~Larman, \emph{Applying {UML} and Patterns: An Introduction to
  Object-Oriented Analysis and Design}.\hskip 1em plus 0.5em minus 0.4em\relax
  Prentice Hall, 1997.

\bibitem{LecunBBH98}
Y.~LeCun, L.~Bottou, Y.~Bengio, P.~Haffner \emph{et~al.}, ``Gradient-based
  learning applied to document recognition,'' \emph{Proceedings of the IEEE},
  vol.~86, no.~11, pp. 2278--2324, 1998.

\bibitem{Selinger03}
P.~Selinger, ``Potrace: a polygon-based tracing algorithm,''
  \url{http://potrace.sourceforge.net/potrace.pdf}, 2003.

\bibitem{zhang:2015}
X.~Zhang, Y.~Sugano, M.~Fritz, and A.~Bulling, ``Appearance-based gaze
  estimation in the wild,'' in \emph{Proceedings of the IEEE conference on
  computer vision and pattern recognition}, 2015, pp. 4511--4520.

\bibitem{wood:2016}
E.~Wood, T.~Baltru{\v{s}}aitis, L.-P. Morency, P.~Robinson, and A.~Bulling,
  ``Learning an appearance-based gaze estimator from one million synthesised
  images,'' in \emph{Proceedings of the Ninth Biennial ACM Symposium on Eye
  Tracking Research \& Applications}, 2016, pp. 131--138.

\bibitem{replicationPkg}
V.~Riccio, N.~Humbatova, G.~Jahangirova, and P.~Tonella, ``Replication package
  for deepmetis,'' \url{https://github.com/testingautomated-usi/deepmetis},
  2021.

\bibitem{DeJong04}
E.~D. de~Jong, ``The incremental pareto-coevolution archive,'' in \emph{Genetic
  and Evolutionary Computation -- GECCO 2004}, K.~Deb, Ed.\hskip 1em plus 0.5em
  minus 0.4em\relax Berlin, Heidelberg: Springer Berlin Heidelberg, 2004, pp.
  525--536.

\bibitem{Mouret15}
J.-B. Mouret and J.~Clune, ``Illuminating search spaces by mapping elites,''
  2015.

\bibitem{unityEyesModel}
``An implementation of a multimodal cnn for appearance-based gaze estimation.''
  \url{https://github.com/dlsuroviki/UnityEyesModel}, 2020.

\bibitem{Byun2019}
\BIBentryALTinterwordspacing
T.~Byun, V.~Sharma, A.~Vijayakumar, S.~Rayadurgam, and D.~Cofer, ``Input
  prioritization for testing neural networks,'' in \emph{2019 IEEE
  International Conference On Artificial Intelligence Testing (AITest)}.\hskip
  1em plus 0.5em minus 0.4em\relax IEEE, 2019, pp. 63--70. [Online]. Available:
  \url{https://doi.org/10.1109/AITest.2019.000-6}
\BIBentrySTDinterwordspacing

\bibitem{Dola21}
S.~Dola, M.~B. Dwyer, and M.~L. Soffa, ``Distribution-aware testing of neural
  networks using generative models,'' \emph{arXiv preprint arXiv:2102.13602},
  2021.

\bibitem{ZhangZZ0K18}
M.~Zhang, Y.~Zhang, L.~Zhang, C.~Liu, and S.~Khurshid, ``Deeproad: Gan-based
  metamorphic testing and input validation framework for autonomous driving
  systems,'' in \emph{Proceedings of the 33rd {ACM/IEEE} International
  Conference on Automated Software Engineering, {ASE}}, 2018, pp. 132--142.

\bibitem{LeeISSTA20}
\BIBentryALTinterwordspacing
S.~Lee, S.~Cha, D.~Lee, and H.~Oh, ``Effective white-box testing of deep neural
  networks with adaptive neuron-selection strategy,'' in \emph{Proceedings of
  the 29th ACM SIGSOFT International Symposium on Software Testing and
  Analysis}, ser. ISSTA 2020.\hskip 1em plus 0.5em minus 0.4em\relax New York,
  NY, USA: Association for Computing Machinery, 2020, p. 165–176. [Online].
  Available: \url{https://doi.org/10.1145/3395363.3397346}
\BIBentrySTDinterwordspacing

\bibitem{AbdessalemNBS16}
R.~B. Abdessalem, S.~Nejati, L.~C. Briand, and T.~Stifter, ``Testing advanced
  driver assistance systems using multi-objective search and neural networks,''
  in \emph{Proceedings of the 31st {IEEE/ACM} International Conference on
  Automated Software Engineering, {ASE}}, 2016, pp. 63--74.

\bibitem{AbdessalemPNBS18}
\BIBentryALTinterwordspacing
R.~B. Abdessalem, A.~Panichella, S.~Nejati, L.~C. Briand, and T.~Stifter,
  ``Testing autonomous cars for feature interaction failures using
  many-objective search,'' in \emph{Proceedings of the 33rd ACM/IEEE
  International Conference on Automated Software Engineering}, ser. ASE
  2018.\hskip 1em plus 0.5em minus 0.4em\relax New York, NY, USA: ACM, 2018,
  pp. 143--154. [Online]. Available:
  \url{http://doi.acm.org/10.1145/3238147.3238192}
\BIBentrySTDinterwordspacing

\bibitem{AbdessalemNBS18}
\BIBentryALTinterwordspacing
R.~B. Abdessalem, S.~Nejati, L.~C. Briand, and T.~Stifter, ``Testing
  vision-based control systems using learnable evolutionary algorithms,'' in
  \emph{Proceedings of the 40th International Conference on Software
  Engineering}, ser. ICSE '18.\hskip 1em plus 0.5em minus 0.4em\relax New York,
  NY, USA: ACM, 2018, pp. 1016--1026. [Online]. Available:
  \url{http://doi.acm.org/10.1145/3180155.3180160}
\BIBentrySTDinterwordspacing

\bibitem{GambiMF19}
A.~Gambi, M.~M{\"{u}}ller, and G.~Fraser, ``Automatically testing self-driving
  cars with search-based procedural content generation,'' in \emph{Proceedings
  of the 28th {ACM} {SIGSOFT} International Symposium on Software Testing and
  Analysis, {ISSTA}}, 2019, pp. 318--328.

\bibitem{UdeshiAC18}
\BIBentryALTinterwordspacing
S.~Udeshi, P.~Arora, and S.~Chattopadhyay, ``Automated directed fairness
  testing,'' in \emph{Proceedings of the 33rd ACM/IEEE International Conference
  on Automated Software Engineering}, ser. ASE 2018.\hskip 1em plus 0.5em minus
  0.4em\relax New York, NY, USA: ACM, 2018, pp. 98--108. [Online]. Available:
  \url{http://doi.acm.org/10.1145/3238147.3238165}
\BIBentrySTDinterwordspacing

\bibitem{Ma-ASE-2018}
\BIBentryALTinterwordspacing
L.~Ma, F.~Juefei-Xu, F.~Zhang, J.~Sun, M.~Xue, B.~Li, C.~Chen, T.~Su, L.~Li,
  Y.~Liu, J.~Zhao, and Y.~Wang, ``Deepgauge: Multi-granularity testing criteria
  for deep learning systems,'' in \emph{Proceedings of the 33rd ACM/IEEE
  International Conference on Automated Software Engineering}, ser. ASE
  2018.\hskip 1em plus 0.5em minus 0.4em\relax New York, NY, USA: ACM, 2018,
  pp. 120--131. [Online]. Available:
  \url{http://doi.acm.org/10.1145/3238147.3238202}
\BIBentrySTDinterwordspacing

\bibitem{ZhangICSE20}
X.~Zhang, X.~Xie, L.~Ma, X.~Du, Q.~Hu, Y.~Liu, J.~Zhao, and S.~Meng, ``Towards
  characterizing adversarial defects of deep learning software from the lens of
  uncertainty,'' in \emph{Proceedings of 42nd International Conference on
  Software Engineering}, ser. ICSE ’20.\hskip 1em plus 0.5em minus
  0.4em\relax ACM, 2020, p. 12 pages.

\end{thebibliography}

\end{document}